\documentclass[12pt,preprint]{aastex}
\usepackage[onecolumn,numberedappendix]{emulateapj5}
\usepackage{epsfig}
\tightenlines

\newcommand \lsim{\mathrel{\rlap{\lower4pt\hbox{\hskip1pt$\sim$}}
    \raise1pt\hbox{$<$}}}
\newcommand \gsim{\mathrel{\rlap{\lower4pt\hbox{\hskip1pt$\sim$}}
    \raise1pt\hbox{$>$}}}

\newcommand{\sm}{{M_\odot}}

\newcommand{\beq}{\begin{equation}}
\newcommand{\eeq}{\end{equation}}
\newcommand{\beqa}{\begin{eqnarray}}
\newcommand{\eeqa}{\end{eqnarray}}

\newlength{\figwidth}
\countdef\decade=200
\advance\decade by \year
\countdef\hours=201
\advance\hours by \time
\divide\hours by 60
\countdef\mins=202
\advance\mins by \hours
\multiply\mins by 60
\multiply\hours by 100
\countdef\miltime=203
\advance\miltime by \hours
\advance\miltime by \time
\advance\miltime by -\mins

\begin{document}

\title{A Search for Molecular Gas in the Nucleus of M87\\ and Implications for the Fueling of Supermassive Black Holes}


\author{Jonathan C. Tan}
\affil{Dept. of Astronomy, University of Florida, Gainesville, FL 32611, USA}
\author{Henrik Beuther, Fabian Walter}
\affil{Max-Planck-Institute for Astronomy, 69117 Heidelberg, Germany}
\author{Eric G. Blackman}
\affil{Depts. of Physics and Astronomy, University of Rochester, Rochester, NY, USA}

\begin{abstract}
Supermassive black holes in giant elliptical galaxies are remarkably
faint given their expected accretion rates. This motivates models of
radiatively inefficient accretion, due to either ion-electron thermal
decoupling, generation of outflows that inhibit accretion, or settling
of gas to a gravitationally unstable disk that forms stars in
preference to feeding the black hole. The latter model predicts the
presence of cold molecular gas in a thin disk around the black
hole. Here we report Submillimeter Array observations of the nucleus
of the giant elliptical galaxy M87 that probe 230~GHz continuum and
CO(J=2--1) line emission. Continuum emission is detected from the
nucleus and several knots in the jet, including one that has been
undergoing flaring behavior. We estimate a conservative upper limit on
the mass of molecular gas within $\sim 100\:{\rm pc}$ and $\pm
400\:{\rm km\:s^{-1}}$ line of sight velocity of the central black
hole of $\sim 8\times 10^6 M_\odot$, which includes an allowance for
possible systematic errors associated with subtraction of the
continuum. Ignoring such errors, we have a $3\sigma$ sensitivity to
about $3\times 10^6M_\odot$.  In fact, the continuum-subtracted
spectrum shows weak emission features extending up to 4$\sigma$ above
the RMS dispersion of the line-free channels. These may be artifacts
of the continuum subtraction process. Alternatively, if they are
interpreted as CO emission, then the implied molecular gas mass is
$\sim 5\times10^6 M_\odot$ spread out over a velocity range of
700~$\rm km\:s^{-1}$.  These constraints on molecular gas mass are
close to the predictions of the model of self-gravitating,
star-forming accretion disks fed by Bondi accretion (Tan \& Blackman
2005).
\end{abstract}


\section{Introduction}\label{S:intro}

The most massive black holes in the Universe reside in the centers of
giant elliptical galaxies. The $(3.4\pm1.0)\times 10^9 M_\odot$
(Macchetto et al. 1997) behemoth in M87 in the Virgo Cluster is one of
the nearest ($d=16\pm 1.2$~Mpc, Tonry et al. 2001, for which
1\arcsec=78~pc). Such black holes are thought to have accumulated most
of their mass by accretion of gas from thin disks, which efficiently
radiate about 10\% of the rest mass energy of the gas. These systems
are likely to power the luminous quasars that are common at high
redshifts.

Most black holes we observe today in the local Universe are not
surrounded by very luminous accretion disks. This is surprising
because these black holes are often embedded in gas that should be
able to accrete. In the centers of elliptical galaxies the gas is
typically at temperatures of $kT \simeq 1$~keV, i.e. $T \simeq
10^7$~K, corresponding to sound speeds of 400~$\rm km\:s^{-1}$, and at
particle densities of about $1\:{\rm cm^{-3}}$. These properties can be
measured relatively accurately with high-spatial-resolution X-ray
telescopes.  For example the Chandra X-ray Observatory measures $kT =
0.8$~keV and $n_e = 0.17\:{\rm cm^{-3}}$ in the central kpc of M87
(Di~Matteo et al. 2003). The expected Bondi accretion rate to the
black hole is given approximately by the product of the gas density,
sound speed, and area of a sphere with a radius such that the escape
speed at this distance is the sound speed. For M87 the expected
accretion rate is about $0.04~M_\odot {\rm yr}^{-1}$ and the
luminosity of a thin accretion disk would then be $2\times 10^{44}\:{\rm
ergs\: s^{-1}}$ (Tan \& Blackman 2005, hereafter TB05). However, the
radiative luminosity of the nucleus is observed to be only $10^{42}\:
{\rm ergs \ s^{-1}}$ (Biretta, Stern \& Harris 1991). 

It is possible that the radiative luminosity is only a fraction of the
total accretion luminosity since accretion likely powers the observed
M87 jet. The mechanical luminosity of the observed M87 jet has been
estimated to be as large as $2\times 10^{43}\:{\rm ergs\: s^{-1}}$
(Reynolds et al. 1996) by assuming a minimum pressure in the 5~kpc
radio lobes to obtain the total energy, in combination with
synchrotron spectral aging to obtain a lifetime of $10^6$ years. These
estimates are sensitive to any uncertainties in the aging. Also, it is
not certain that the M87 outflow luminosity has been steady from the
beginning of the radio lobe formation to the present. Nevertheless, if
the jet luminosity were steady and the above estimate correct, then
the discrepancy between a model involving the extraction of 10\% of
the rest mass energy of an accretion flow equal to the present Bondi
Accretion rate and the total output from accretion (radiative +
mechanical) would still be a factor of 10.

To help explain these discrepancies, models of radiatively inefficient
accretion flows (RIAFs) have been proposed. In the simplest variant,
the cooling time of the gas is long compared to the accretion time, so
much of the energy is advected into the black hole (Ichimaru 1977;
Narayan \& Yi 1995; Quataert \& Gruzinov 1999). In other variants, the
inability of the gas to cool leads to reduced accretion rates near the
black hole due to the generation of outflows (Blandford \& Begelman
1999) or convective motions (Quataert \& Gruzinov 2000). A
qualitatively different model allows the gas to cool to form a thin
disk, but prevents accretion of the gas by having it form stars
(TB05). These stars are then on relatively dissipationless orbits
about the black hole. A prediction of this model is the presence of a
star-forming disk inside the Bondi radius of the black hole. Since all
known star formation in the Universe occurs in cold, molecular gas,
that is best traced by the rotational transitions of carbon monoxide
(CO), this model predicts that CO emission should be seen from such a
circumnuclear disk.

In the case of M87, TB05 predicted a molecular gas mass of about
$(1-5) \times 10^6 M_\odot$, assuming the disk is self-regulated to
have a Toomre $Q$ parameter of unity by the energy input from star
formation at a rate equal to the Bondi accretion rate. This star
formation rate is consistent with the observed H$\alpha$ luminosity of
the nuclear disk, although it is possible that the AGN also contributes
significantly to the ionization of this gas. The quoted variation in
this estimate is due to whether the disk is optically thick or thin to
the heating radiation from the stars. Additional systematic
uncertainties at the factor of a few level are caused by uncertainties
in the stellar initial mass function (IMF), radial distribution of the
gas and star formation, and the black hole mass.

The deep potential well of the M87 black hole causes large orbital
speeds in the disk. Given the above mass estimate, these are equal to
$382\pm60 \:{\rm km\:s^{-1}}$ at a distance of 100~pc, about the
extent of the observed H$\alpha$ (Ford et al. 1994) and \ion{O}{2}
disk (Macchetto et al. 1997).  From the latter study, the best fit
value of the inclination angle of the disk's rotation axis to our line
of sight is $51^\circ$, the systemic velocity is 1290~$\rm
km\:s^{-1}$, and the observed FWHM along a slit running through the
center of the disk and 1\arcsec\ in extent is about 700~$\rm
km\:s^{-1}$.

If CO emission emerges from the same region, then one expects a similarly
broad line, which presents a problem for its detection since most
radio telescopes have relatively narrow band passes. The Submillimeter
Array (SMA\footnote{The Submillimeter Array is a joint project between
the Smithsonian Astrophysical Observatory and the Academia Sinica
Institute of Astronomy and Astrophysics, and is funded by the
Smithsonian Institution and the Academia Sinica.}, Ho, Moran, \& Lo
2004) has a large band pass of about 2~GHz, corresponding to 2600~$\rm
km\:s^{-1}$ at the frequency of CO(J=2--1) line emission
(i.e. 230~GHz, 1.3mm). Having significant line-free regions in the
bandpass is important for determining the continuum level, since the
immediate vicinity of the black hole is known to be a strong
($\sim$1.5~Jy) radio continuum source at $\lambda$=1\,mm. For these
reasons, we used the SMA to perform a search for molecular gas in the
nuclear disk of M87. 

\subsection{Previous searches for CO in M87}

Braine \& Wiklind (1993) searched for CO in M87 with the IRAM 30m
single-dish telescope, with a beam size of 13\arcsec(=1.0~kpc) for the
CO(2-1) search at 230~GHz and 21\arcsec(=1.6~kpc) for the CO(1-0)
search at 115~GHz. Thus these observations did not resolve the $\rm
H\alpha$ circumnuclear disk. Another limitation was the relatively
narrow bandpass, corresponding to $700\:{\rm km\:s^{-1}}$ for CO(2-1)
and $1300\:{\rm km\:s^{-1}}$ for CO(1-0). Braine \& Wiklind (1993)
derived an upper limit to the molecular gas mass of $3\times
10^6\:M_\odot$ from their CO(2--1) data (based on a $3\sigma$
uncertainty of $0.9~{\rm K\:km\:s^{-1}}$ over the assumed 500~$\rm
km\:s^{-1}$ line width, a J=2-1 to J=1-0 line ratio of unity, a
standard CO to $\rm H_2$ conversion factor of $N({\rm H_2})/I_{\rm
CO(1-0)} =2.3 \times 10^{20}\:{\rm mol\:cm^{-2} / (K\:km s^{-1})}$ for
optically thick emission at temperatures $\sim10-20$~K, and a
geometric area of $\sim \pi (500~{\rm pc})^2$) and a limit of $1\times
10^7\:M_\odot$ from their CO(1-0) data. The limit from CO(2--1) is not
valid for a line that is broader than about $500\:{\rm
km\:s^{-1}}$. Braine \& Wiklind (1993) also derived more stringent
limits assuming optically thin CO emission, although this would
require $N({\rm H_2}) \lesssim 10^{21} \:{\rm cm^{-2}}$ for typical
molecular cloud conditions, which would not be expected if the gas was
organized into self-gravitating molecular clouds, as in the model of
TB05.

Combes, Young \& Bureau (2007) also searched for CO(1-0) and CO(2-1)
emission from the central region of M87 with the IRAM~30m single-dish
telescope. From the lack of detected CO(1-0) emission in their
23\arcsec (1.8~kpc) FWHM beam and 1300~$\rm km\:s^{-1}$ bandpass, they
estimate a $3\sigma$ upper limit of $1.1\times 10^7\:M_\odot$ of
molecular gas (assuming $N({\rm H_2})/I_{\rm CO(1-0)} =3 \times
10^{20}\:{\rm mol\:cm^{-2} / (K\:km s^{-1})}$, a 300~$\rm km\:s^{-1}$
line width, and a geometric area of $\sim \pi (900{\rm pc})^2$). Using
similar observations of CO(1-0) with the IRAM 30m telescope, Salom\'e
\& Combes (2008) reported an upper mass limit of $7.7\times
10^{6}\:M_\odot$ ($1\sigma$ corrected to our adopted distance of
16~Mpc) evaluated assuming a 300~$\rm km\:s^{-1}$ line width. We will
typically consider $3\sigma$ upper limits, which for their results is
$2.2\times 10^7\:M_\odot$.

\section{Observations}\label{S:observations}

The M87 nucleus was observed with the SMA at 230~GHz in the extended
configuration with 7 antennas in the array during the night to
February 10th 2006. The primary beam at the given frequency was $\sim
55''$, and the unprojected baselines ranged between 28 and 226\,m.
The phase center of the observations was R.A. (J2000.0)
12$^h$30$^m$49.$''$4 and Dec. (J2000.0) 12$^{\circ}$23$'$28.$''$0. We
had excellent weather conditions with measured zenith opacities
$\tau(\rm{225GHz})$ between 0.03 and 0.06.

Bandpass calibration was done with the quasar 3C273. Based on the
spectrum taken toward the additionally observed quasar 3C279, we
estimate the bandpass accuracy to be good to approximately 1\%, which
was only possible because of the excellent weather conditions.
The flux calibration was performed with measurements of Callisto. We
estimate the absolute flux uncertainties to be within 20\%. Phase and
amplitude calibration were done via frequent observations of the quasar
3C273, about 10.3$^{\circ}$ from the phase center. The absolute
positional accuracy is estimated to be good to within $0.1''$.  The
receiver operated in a double-sideband mode with an IF band of
4-6\,GHz so that the upper and lower sidebands were separated by
10\,GHz. The central frequency of the upper sideband was centered at
the CO(2--1) line at 230.538\,GHz, the $v_{\rm{lsr}}$ was set to
1200\,$\rm km\:s^{-1}$. The correlator had a bandwidth of 2\,GHz and the channel
spacing was 0.8125\,MHz.  Measured double-sideband system temperatures
corrected to the top of the atmosphere were between 70 and 200\,K,
depending on the zenith opacity and the elevation of the source.  The
initial flagging and calibration was done with the IDL superset MIR
originally developed for the Owens Valley Radio Observatory
(Scoville et al. 1993) and adapted for the SMA\footnote{The MIR cookbook
  by Charlie Qi can be found at
  http://cfa-www.harvard.edu/$\sim$cqi/mircook.html.}. The imaging and
data analysis were conducted in MIRIAD (Sault, Teuben, \& Wright 1995).

Since the upper limit of the CO(2--1) line intensity is $\la 5$\%
of the mm continuum, we averaged the whole upper sideband bandpass
to produce our mm continuum image. This small line-to-continuum ratio
made the CO line imaging and the connected continuum subtraction to
the data a very difficult task. We tried various different approaches
for the continuum subtraction using the MIRIAD and the AIPS software packages
to verify consistent results. The final dataset was produced in MIRIAD
using the task UVLIN and setting a spectral window between 700 and
1700\,$\rm km\:s^{-1}$ to ensure exclusion of any potential line
contamination in the continuum subtraction fits. The final synthesized
beams for the continuum and line datasets are $1.2''\times 0.8''$. Our
sensitivity was dynamic-range limited by the side-lobes of the
strongest emission peaks. 
The $1\sigma$ rms dispersion of the continuum data in regions of the
spectrum expected to be free of CO(2--1) line emission and
averaged over 100\,$\rm km\:s^{-1}$ channels is 6.8\,$\rm
mJy\:beam^{-1}$. Since our shortest baseline was 28\,m, this dataset
filters out all emission on scales larger than $\sim 12''$.

\section{Results}

\begin{figure}
\plotone{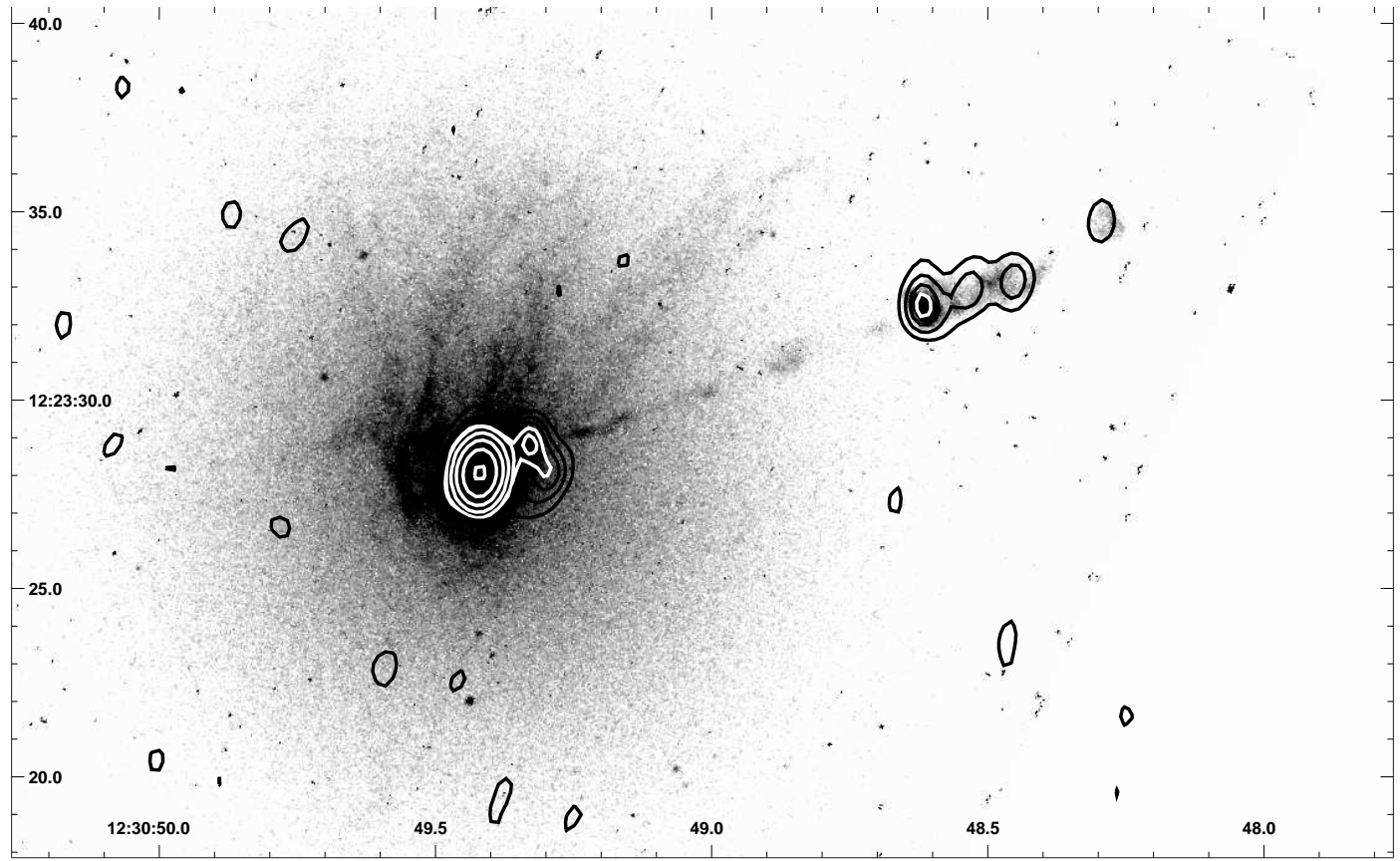}
\caption{\label{fig:continuum} 230~GHz continuum image of the nucleus
and jet of M87 (contours), superposed with the optical HST
image from Ford et al. (1994).
The contouring of the 230\,GHz data is from 18 to 90\,mJy beam$^{-1}$
in 18\,mJy beam$^{-1}$ steps, then each contour level is at double the
intensity of the previous. White and black contours are used for
display purposes and are part of the same sequence.
}
\end{figure}

\begin{deluxetable}{cccc} 
\tablecaption{Nucleus and Jet Knot Fluxes\label{tab:knots}}
\tablewidth{0pt}
\tablehead{
\colhead{Component} & \colhead{1.3~mm Flux (mJy)} & \colhead{$\alpha_{\rm ir-mm}$} & \colhead{$\alpha_{\rm mm}$}
}
\startdata
Nucleus & $1770\pm350$ & $0.97\pm0.04$ & $0.06\pm0.26$\\
HST1/2 & $180\pm36$ & n/a & n/a\\
A & $101\pm20$ & $0.58\pm0.04$ & $0.55\pm0.26$\\
B & $98\pm20$ & $0.57\pm0.04$ & $1.28\pm0.26$\\
C & $23\pm5$ & $0.36\pm0.05$ & $2.4\pm0.26$\\
\enddata
\end{deluxetable}

\subsection{230~GHz Continuum Image of Nucleus and Jet}

The 230~GHz continuum image of the M87 nuclear region is shown in
Fig.~\ref{fig:continuum}, together with the 
H$\alpha$ image from HST (Ford et al. 1994). At 230~GHz, the nucleus is
detected at a position (0.3\arcsec, 0.0\arcsec) relative to the phase
center (\S \ref{S:observations}). The peak and integrated fluxes of
the nucleus are 1520\,$\rm mJy\:beam^{-1}$ and 1770\,mJy, respectively.
The HST image was shifted by $1.06\arcsec$ (consistent with HST
pointing uncertainties) to the south-west to align the nucleus and
jet axis with the sub-mm data.

\begin{figure}
\includegraphics[angle=-90,width=6in]{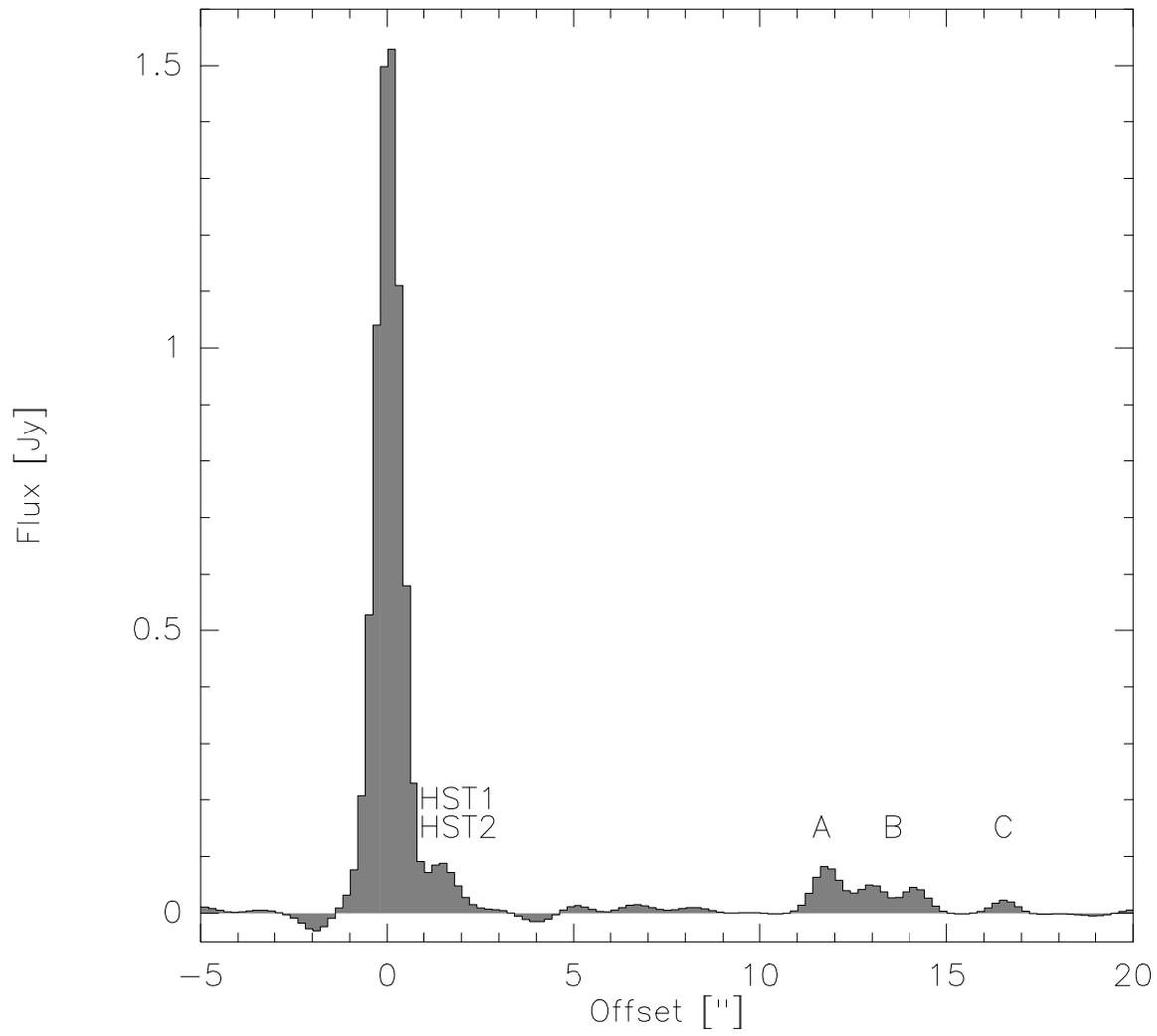}
\caption{\label{fig:strip} 230~GHz continuum emission, integrated
across a 4\arcsec wide strip, as a function of angular distance from
the nucleus along the jet axis.}
\end{figure}

Figure~\ref{fig:continuum} also shows extended emission from the jet
at about 1-2\arcsec\ (Knots HST-1 and HST-2) (Biretta, Sparks, \&
Macchetto 1999) and at 11-18\arcsec\ (Knots A, B, C) away from the
nucleus (see also Fig.~1 of Perlman et al. 2001).
Figure~\ref{fig:strip} shows a profile of this emission along the jet
axis and Table~\ref{tab:knots} lists the knot fluxes. Note that knots
A and B are really one physically-connected structure, and that knot B
has two separate sub-peaks. When deriving fluxes for knots A and B we
define their boundary at the location of the minimum flux between the
knots, at about 12.5\arcsec\ from the nuclear emission peak. We assign
20\% uncertainties to the knot fluxes (\S\ref{S:observations}).

We compare our nuclear and knot fluxes to those measured in the
infrared at 10.8~$\rm \mu m$ ($2.78\times 10^{13}$~Hz) by Perlman et
al. (2001, their Table~1) and those measured at 3.4~mm (88.26~GHz) by
Despringre et al. (1996, their \S3.2, assuming 20\% flux
uncertainties): assuming a power law spectrum, $F_\nu \propto
\nu^{-\alpha}$, we derive spectral indices $\alpha_{\rm ir-mm}$ and
$\alpha_{\rm mm}$, respectively (Table~\ref{tab:knots}). Care must be
taken in the interpretation of these spectral indices since the
nucleus and jet components may be variable: there is a 4.8 year time
baseline between our 1.3~mm observations and the 10.8~$\rm \mu m$
observations and about a 12.5 year time baseline for our comparison
with the 3.4~mm observations. For example, Steppe et al. (1988) found
the nuclues of M87 is variable by about 1~Jy~$\rm yr^{-1}$ at 90~GHz.
Braine \& Wiklind (1993) measured a 1.3~mm nuclear ($<13\arcsec$) flux
of 1.0~Jy from their observations taken in Nov. 1991, a factor of 0.56
smaller than the value measured in this paper.

\subsubsection{Nucleus: Thermal Emission?}

Our value of $\alpha_{\rm ir-mm}\simeq 0.97\pm0.04$ for the nucleus is
similar to the value of 0.88 predicted by the continuous injection
synchrotron model (Heavens \& Meisenheimer 1987) that was fitted to
radio, optical and x-ray data by Sparks, Biretta, \& Macchetto (1996)
and Marshall et al. (2002). The spectral index we derive from
comparison to the 3.4~mm data of Despringre et al. (1996) is
$\alpha_{\rm mm}=0.06\pm0.26$. This is similar to the index of
$0.10\pm0.02$ derived from radio (1.47, 4.89, 15.0~GHz) data by
Biretta et al. (1991), but shallower than the index derived by
comparing the 3.4~mm data to the radio data.

\begin{figure}[h]
\begin{center}
\includegraphics[angle=0,width=6in]{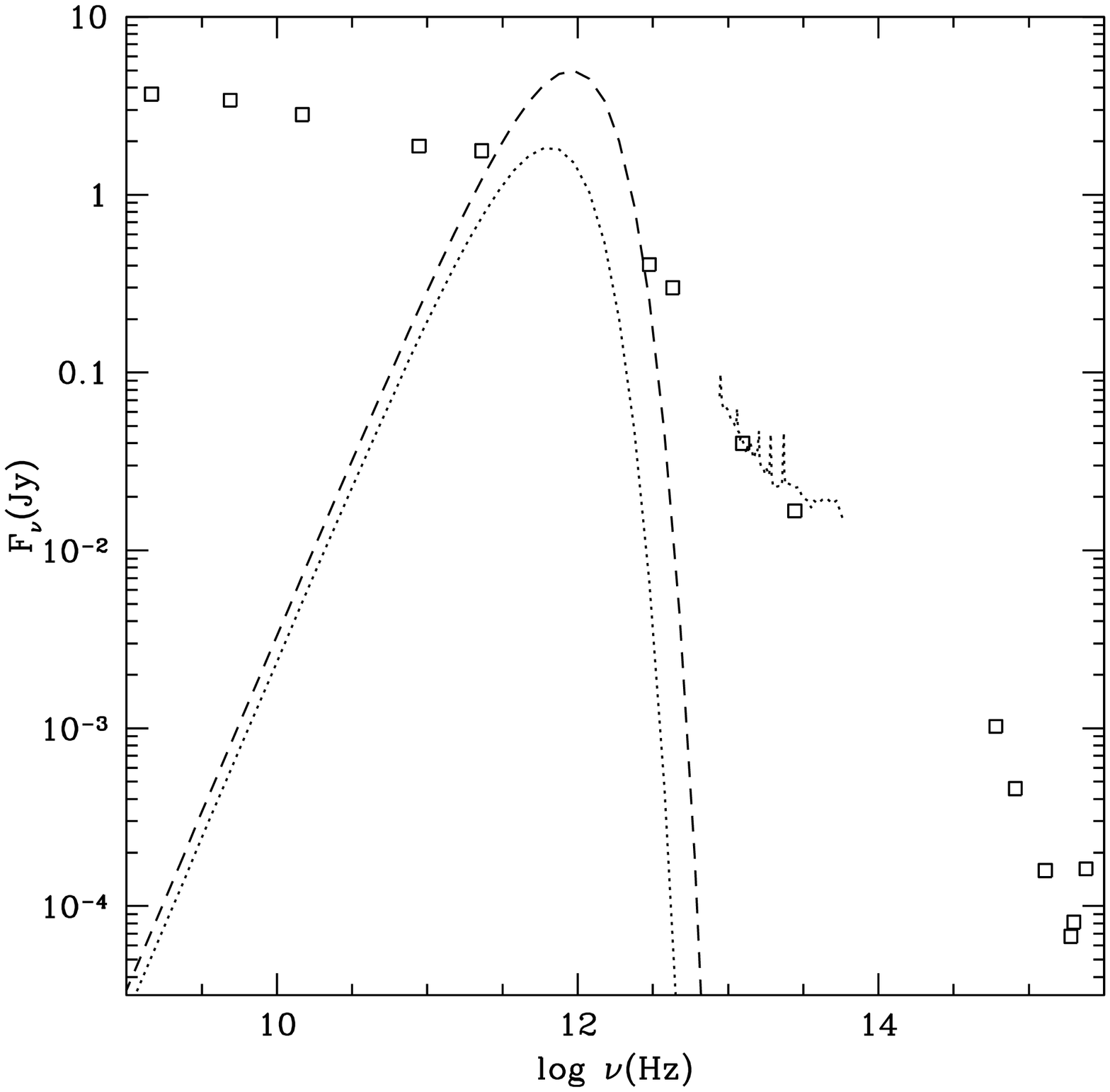}
\end{center}
\caption{\label{fig:flux} Continuum spectrum of the M87 nucleus. Data
points are (from low to high frequency): 1.5, 5, 15~GHz (Biretta et
al. 1991); 89~GHz (Despringre et al. 1996); 230~GHz (this paper); IRAS
100~$\rm \mu m$, Spitzer-MIPS 70 and 24~$\rm \mu m$ and Spitzer-IRS
(5-35~$\rm \mu m$, dotted line) (Perlman et al. 2007); Gemini~10~$\rm
\mu m$ (Perlman et al. 2001); optical and UV data (Sparks et
al. 1996). The dashed and dotted lines show models of thermal dust
emission from TB05, assuming luminosities of
$5.9,1.6\times 10^8\:L_\odot$ and temperatures of 15.5, 11.1~K,
respectively. Note, these are simple, single-temperature, blackbody
models that assume the entire luminosity due to star formation at
rates equal to 0.15, 0.036~$M_\odot\:{\rm yr^{-1}}$ with Salpeter
initial mass function down to 0.1~$M_\odot$ is reprocessed by dust in
a 100~pc radius disk.}
\end{figure}

We show the continuum spectrum of the nucleus in
Fig.~\ref{fig:flux}. The star-forming accretion disk model of TB05
predicts that there should be thermal emission from dust in the disk
and the simple fiducial models of this emission are also shown in
Fig.~\ref{fig:flux}. These models assumed full reprocessing of the
luminosity produced by stars forming (with Salpeter initial mass
function down to 0.1~$M_\odot$) at rates equal to the range of
predicted Bondi accretion rates (0.036-0.15~$M_\odot\:{\rm
yr^{-1}}$). The luminosity was assumed to be radiated with a single
temperature blackbody spectrum from a disk with radius of 100~pc. In
reality, only a fraction of any stellar luminosity is likely to be
reprocessed by dust and there will be a range of temperatures
extending to hotter values. Thus the flux in the Rayleigh-Jeans part
of the spectrum (e.g. at 230~GHz) predicted by these models will tend
to be an overestimate. More accurate models of the thermal emission
are in the process of being developed (Tan, in prep.). 

In any case, our observed flux at 230~GHz is greater than the amount
predicted by the TB05 models. The relatively shallow spectral index
that we find from a comparison with the 3.4~mm data of Despringre et
al. (1996) could be explained by the presence of a star-formation
induced thermal component. Alternatively, it could be due to source
variability or additional uncertainties in the measured fluxes. Note,
Despringre et al. (1996) found extended (12.7\arcsec\ by 6.5\arcsec)
3.4~mm emission around the nucleus with a flux of 1.4~Jy. Although
this emission is aligned in a direction perpendicular to the jet, its
scale is somewhat larger than that expected by a Bondi-fed accretion
disk. Our 230~GHz interferometric observations do not show any
evidence for such an extended component, although they lose
sensitivity for scales $\gtrsim 12\arcsec$.

Perlman et al. (2007) have found evidence for a thermal dust component
in the nucleus of M87 based on Spitzer infrared data from 5 to 70~$\rm
\mu m$. They fit models with temperatures in the range 45 to 65~K (for
which the luminosities are $8.1\times 10^{5} - 7.5\times
10^4\:L_\odot$ and the disk masses are $2.2\times 10^4 - 2.3\times
10^3\:M_\odot$, respectively). Their data do not exclude the possible
presence of additional cooler dust components.

\subsubsection{Jet Knots}

Our values of $\alpha_{\rm ir-mm}\simeq 0.6$ for the jet knots A and B
are very similar to the value of 0.68 predicted by the synchrotron
models of Jaffe \& Perola (1973), Kardashev (1962) and Pacholczyk
(1970) presented by Marshall et al. (2002) and Perlman et al. (2001)
based on the radio and optical data presented by Sparks, Biretta, \&
Macchetto (1996). Our value of $\alpha_{\rm ir-mm}\simeq 0.36\pm0.05$
for knot C is somewhat shallower than predicted, but it is the weakest
of the knot components that we detect and we may have underestimated
the uncertainty in its flux. From the comparison with the 3.4~mm data,
we find a similar spectral index as the synchrotron models for knot A,
$\simeq0.6$, but much steeper indices $\gtrsim 1-2$ for knots B and
C. The knot components were not well resolved by Despringre et
al. (1996) so we evaluate the spectral index derived from the total
flux of A, B and C: in this case we find $\alpha_{\rm mm}=1.2\pm0.26$,
i.e. still significantly steeper than the synchrotron models. This
discrepancy could be due to variability in the knot fluxes or
additional systematic uncertainties in the flux measurements.

The peak and integrated fluxes of the inner sub-mm knot (sum of HST-1,
known from optical and UV observations to be 0.85\arcsec\ from the
nucleus, and HST-2) are 101\,$\rm mJy\:beam^{-1}$ and 180\,mJy,
respectively. Knot HST-1 has been undergoing a flare in the last
several years: its 2~keV x-ray flux increased by about a factor of 50
(Harris et al. 2006), peaking at a date of about April 2005 and
declining to about half its peak value by August 2005.  The rising
part of the flare is also seen in UV (220~nm) and radio (15~GHz) data
(Harris et al. 2006). It is likely that our measurement of the 230~GHz
continuum in February 2006 includes a contribution from the declining
part of this flare.

\subsection{Search for CO(J=2-1) Emission}

\begin{figure}
\includegraphics[angle=-90,width=6in]{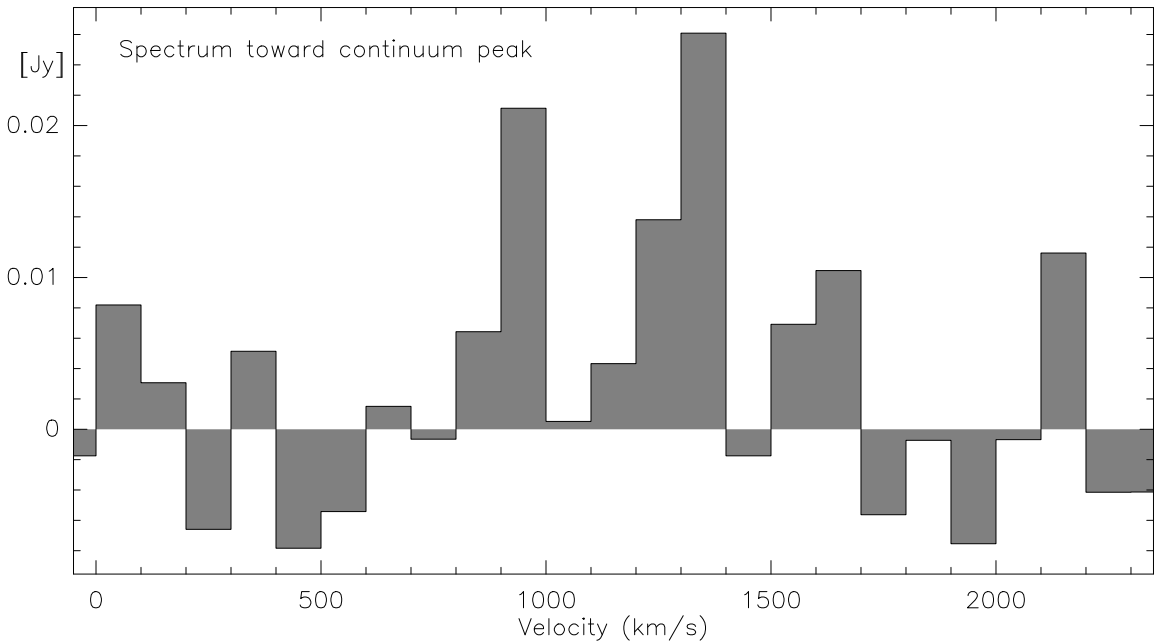}
\caption{\label{fig:spectrum} Continuum-subtracted spectrum at the
positions of the nucleus of M87. The spectrum was extracted toward the
continuum peak position as shown in Figure 1. The systemic velocity of
the M87 black hole is estimated to be at 1290~$\rm km\:s^{-1}$
(Macchetto et al. 1997). The regions of the spectrum expected to be
line-free are 0--700~$\rm km\:s^{-1}$ and 1700--2300~$\rm km\:s^{-1}$,
i.e. at least 400~$\rm km\:s^{-1}$ from the systemic velocity, for
which the rms dispersion is 6.8~mJy per 100~$\rm km\:s^{-1}$
channel. A few channels in the intervening region show weak emission,
but at a level too small to be considered a significant detection (see
text).}
\end{figure}

The continuum-subtracted spectrum extracted at the position of the
continuum peak is shown in Fig.~\ref{fig:spectrum}. Weak ($\sim
3-4\sigma$) emission features are seen in individual channels at 950
and 1350~$\rm km\:s^{-1}$, with 21 and 26~mJy, respectively. Note the
dispersion of the line-free regions of the spectrum is
6.8\,mJy in a 100~$\rm km\:s^{-1}$ channel. 

The continuum level is at 1.5~Jy, so the uncertainties in line
detection are likely to be dominated by the systematic uncertainties
associated with the continuum subtraction. The fact that the emission
features are not in adjacent velocity channels tends to argue in favor
of their origin from systematic errors in the continuum subtraction
process, although if the gas were distributed evenly in an annulus
from 50 to 100~pc from the black hole, a double-peaked profile
separated by about 500~${\rm km\:s^{-1}}$ would be expected
(Fig.~\ref{fig:theory}). If the molecular gas is further organized
into spiral arms or a small number of discrete clouds, this could also
lead to separated emission peaks.

\subsubsection{Applicability of Standard $\rm CO-H_2$ Conversion Factors}

The intensity of the CO line depends on the mass of molecular gas in
the disk and its excitation conditions. If the gas is optically thick
in the CO J=1--0 line then the column of gas is related to the
intensity of the line via $N({\rm H_2}) = 1.8\pm 0.3 \times 10^{20}
(I_{\rm CO,J=1-0} / {\rm K\ km\:s^{-1}} ) {\rm cm^{-2}}$ (Dame,
Hartmann, \& Thaddeus 2001), based on observations of molecular clouds
in the Milky Way.

The applicability of this standard conversion factor to the nucleus of
M87 is uncertain. There is some evidence that the conversion factor
varies in different galaxies: e.g. in the center of M51, Israel ,
Tilanus \& Baas (2006) find the CO-to-$\rm H_2$ ratio is
about a factor of 4 lower than the above Galactic value, i.e. four
times less $\rm H_2$ for a given CO line intensity. 

X-ray ionization from the nucleus may be an important factor in
determining CO abundance (Maloney, Hollenbach, \& Tielens 1996). The
structure of these X-ray dissociation regions is controlled by the
ratio of the x-ray flux to the gas density, described in the models of
Maloney et al. (1996) by the dimensionless effective (attenuated)
X-ray ionization parameter, $\xi_{\rm eff}=1.26\times 10^{-2} F_x
n_3^{-1} N_{22}^{-0.9}$, where $F_x$ is the normally incident X-ray
flux in $\rm ergs\:cm^{-2}\:s^{-1}$, $n=1000 n_3\:{\rm cm^{-3}}$ is
the number density of H nuclei, and $N=10^{22}N_{22}\:{\rm cm^{-2}}$
is the absorbing column density between the nucleus and the gas being
considered. The M87 nucleus has $L_x\simeq 10^{40}\:{\rm
ergs\:s^{-1}}$ (Wilson \& Yang 2002) so that $F_x=8.4\times 10^{-3}
(r/100\:{\rm pc})^{-2}\:{\rm ergs\:cm^{-2}\:s^{-1}}$. If there is
$10^{6}\:M_\odot$ of gas in a thin disk with outer radius of 100~pc
(similar to the fiducial estimates of TB05 and the sensitivities being
probed by our SMA observations), then the column density normal to the
disk would be $N_{22}=0.3$. If the vertical scale-height is 10~pc,
i.e. for a thin disk aspect ratio of 0.1, then the mean density in the
disk is $n=100\:{\rm cm^{-3}}$ and the column density through 100~pc
of the disk is $N_{22}=3$. For a cloud that is overdense by a factor
of 10 from the mean, i.e. $n_3=1$, we therefore estimate $\xi_{\rm
eff}\simeq 1.1\times 10^{-4}$. The results of Maloney et al. (1996,
their Fig.~3b) indicate the gas would be essentially fully molecular,
with standard CO abundances,
and a temperature of about $20$~K. If $\xi_{\rm eff}$ were higher by a
factor of 10, then the CO abundance is predicted to be reduced by a
similar factor. We conclude that if the observed 100~pc scale
circumnuclear disk in M87 contains a gas mass of $\sim 10^6\sm$,
allowing gravitational instabilities to form over-dense clouds (TB05),
then the present nuclear X-ray fluxes are small enough to allow
formation of molecular clouds of similar CO content and temperature to
giant molecular clouds seen in the Milky Way. As disk material
approaches the black hole, the increasing X-ray flux will eventually
cause the gas to transition to a fully atomic state.

Radiative pumping of lower excitation CO rotational states may also be
important in affecting the observed intensity of CO(J=2--1) emission
(Maloney, Begelman, \& Rees 1994). The effectiveness of radiative
excitation compared to collisional excitation depends on the incident
mm flux from the AGN and the gas density. For an observed 230~GHz flux
of 1.8~Jy, the flux density at distance $r$ from the nucleus is
$4.6\times 10^{-13} (r/100\:{\rm pc})^{-2}\:{\rm
ergs\:cm^{-2}\:s^{-1}\:Hz^{-1}}$ and the corresponding radiative
excitation rate for $J=1\rightarrow2$ is $7.1\times 10^{-8}\:{\rm
s^{-1}}$. At 20~K the collisional rate coefficient with $\rm H_2$ is
$\gamma_{12}=3.1\times 10^{-11} \:{\rm cm^{-3}}$ with a weak temperature
dependence. The rates of radiative and collisional excitation will be
equal at densities of $2\times 10^3\:{\rm cm^{-3}}$. Thus molecular
gas at densities of $n_3\sim1$ will have its CO excitation temperature
raised above the kinetic temperature by factors of order unity due to
the nuclear radiation source.


With these caveats in mind, in the following, for simplicity, we
proceed using the standard Galactic value to derive $\rm H_2$ column
densities. We also assume that the molecular gas is thermalized up to
the J=2--1 transition, i.e. that the ratio in surface brightness
between the 1--0 and 2--1 transition is unity, as is observed in the
center of our Galaxy and others (Braine \& Combes 1992; Sawada et
al. 2001).

\subsubsection{Estimates of Molecular Gas Mass}

The continuum-subtracted spectrum in the range 700 to 1700~$\rm
km\:s^{-1}$ has a total integrated intensity of 250~$\rm
K\:km\:s^{-1}$, using a Planck conversion factor of 28.3~$\rm
K\:Jy^{-1}$. Assuming independent $1\sigma$ errors of 6.8~mJy per
100~$\rm km\:s^{-1}$ channel, the $1\sigma$ error in this total is
about 25\%. However, as discussed above, the true uncertainties are
likely to be correlated systematic errors due to continuum
subtraction. If the errors are 6.8~mJy per 100~$\rm km\:s^{-1}$
channel and perfectly correlated, (i.e. in the same direction for all
channels), then the uncertainty in the total intensity is about 80\%,
i.e. it could be consistent with zero.

Assuming the intensity in the J=1--0 line is the same as that we
estimate in the J=2--1 line, then $N({\rm H_2}) = 4.5\pm (1.1,3.6)
\times 10^{22}\:{\rm cm^{-2}}$, i.e. $\Sigma = 0.21\pm (0.05,0.17)
{\rm g \ cm^{-2}}$, including He, where the first error assumes
independent velocity channel errors and the second assumes perfectly
correlated velocity channel errors. If this is the mean surface
density over the 1\arcsec\ wide beam (geometric area $=\pi (39{\rm
pc})^2$), then the total gas mass in this beam is
$4.9\pm(1.2,3.9)\times 10^6 M_\odot$. Note that because of the disk's
inclination to our line of sight, this area would probe regions that
are up to 62~pc from the black hole.

If we treat the errors in the 100~$\rm km\:s^{-1}$ channels
independently, and integrate over an assumed line width of 500~$\rm
km\:s^{-1}$ (following Braine \& Wiklind 1993), then our $3\sigma$
sensitivity is $2.5\times 10^6\:M_\odot$, similar to the value they
derived, but for a region that has an area 170 times smaller and a for
a band pass that is almost 4 times wider.

Another method of estimating an upper limit to the presence of
molecular gas within a certain velocity interval comes by assuming
that three adjoining channels are each required to have a $>3\sigma$
detection. This would correspond to an intensity of 173~$\rm
K\:km\:s^{-1}$ and a mass of $3.3 \times 10^6 M_\odot$ within a
velocity interval of 300~$\rm km\:s^{-1}$. Requiring $>3\sigma$
detections in all channels over a 700~$\rm km\:s^{-1}$ velocity
interval corresponds to a mass sensitivity of $8\times 10^6\:M_\odot$.

Given the potential presence of correlated systematic errors,
our results do not represent a definitive detection of
CO emission from molecular gas. The data suggest an upper limit on the
mass of molecular gas that is present within a velocity range $\sim
700\:{\rm km\:s^{-1}}$ about the systemic velocity of $\sim 8\times
10^6\:M_\odot$.

\begin{figure}[h]
\begin{center}
\includegraphics[angle=0,width=6in]{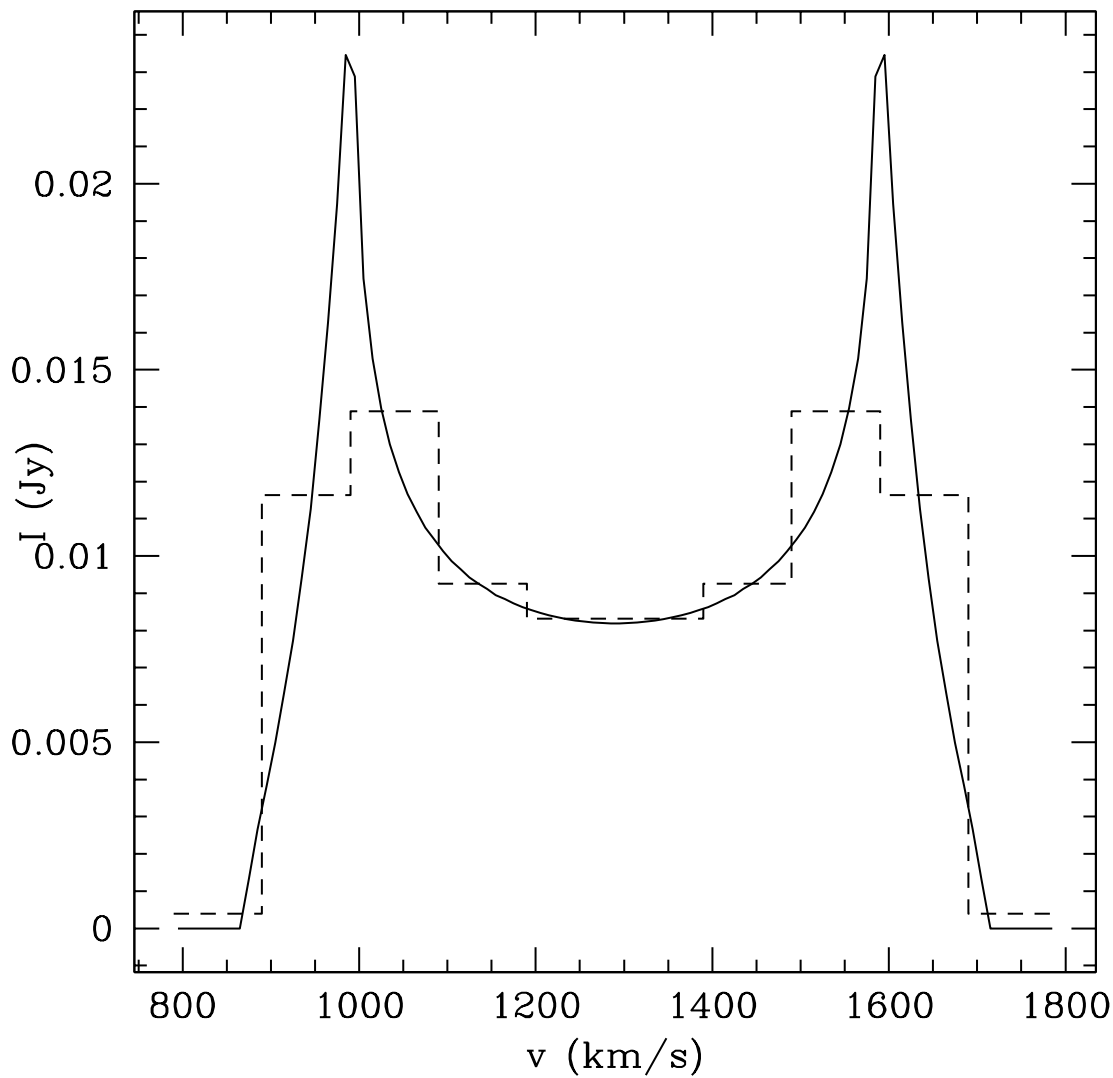}
\end{center}
\caption{\label{fig:theory} Theoretical (zero noise) spectrum (solid line) of
CO(2--1) emission from a thin circumnuclear disk around a $3.4\times
10^9\:M_\odot$ black hole in M87 with $4.9\times 10^6\:M_\odot$ of gas
uniformly distributed from 50 to 100~pc and with the disk inclined by
51$^\circ$ to the line of sight and a central velocity of 1290~$\rm
km\:s^{-1}$ (Macchetto et al. 1997). The dashed line shows the
spectrum averaged into 100~$\rm km\:s^{-1}$ wide channels.
}
\end{figure}

To illustrate the nature of a potential emission line of CO, in
Fig.~\ref{fig:theory} we show the expected line profile from a thin
circumnuclear disk around a $3.4\times 10^9\:M_\odot$ black hole in
M87 with $4.9\times 10^6\:M_\odot$ of gas uniformly distributed from
50 to 100~pc and with the disk inclined by 51$^\circ$ to the line of
sight and a central velocity of 1290~$\rm km\:s^{-1}$ (Macchetto et
al. 1997).

For a disk fed by Bondi accretion, the size of the disk is limited to
within the Bondi radius (about 110~pc, 1.4\arcsec for M87). This is about
the extent of the observed H$\alpha$ disk in M87 (Ford et al. 1994).
The observed velocity gradient of the H$\alpha$ disk is along a
position angle approximately perpendicular to the jet axis. However,
we do not find evidence for a velocity gradient in the potential
CO emission features in our data.

The possibility of seeing CO absorption features against an AGN sub-mm
continuum source has been discussed by, for example, Barvainis \&
Antonucci (1994), Maloney, Begelman, \& Rees (1994), Wiklind \& Combes
(1996), and Liszt \& Lucas (1998). We do not expect local absorption
of the nucleus from the 100~pc scale circumnuclear disk because of its
large inclination angle to our line of sight. Absorption due to CO
distributed on larger scales with M87 would be expected to produce a
relatively narrow ($\lesssim 300\:{\rm km\:s^{-1}}$) feature centered
around the systemic velocity of the galaxy (1290~$\rm
km\:s^{-1}$). This could explain the relative dearth of CO emission
that we see from 1000-1300~$\rm km\:s^{-1}$ compared to the
surrounding channels. Estimating a reduction in flux of $\sim
0.015$~Jy relative to the continuum level of 1.5~Jy, implies
$\tau_{\rm CO(2-1)}=0.01$. Assuming this is constant over a 300~$\rm
km\:s^{-1}$ velocity range and using the conversion factor derived by
Wiklind \& Combes (1996) (for 16~K molecular gas), we estimate a total
absorbing column $N({\rm CO})=1.8\times 10^{16}\:{\rm cm^{-2}}$ so
that $N({\rm H_2})\sim 1.4\times 10^{20}\:{\rm cm^{-2}}$. This is
close to, but slightly lower than, the upper limits measured by Braine
\& Wiklind (1993). Thus it is possible that absorption by a larger
scale distribution of CO, if present in M87, could have a significant
influence on the appearance of any CO emission spectrum from the
circumnuclear disk, reducing the observed flux near the systemic
velocity and thus causing its profile to assume a more double-peaked
appearance.

\section{Conclusions}\label{S:discussion}

We have observed 230~GHz continuum emission from the nucleus and jet
of M87. The measured fluxes are generally consistent with those
expected from synchrotron models of these components, although it is
possible that thermal dust emission makes some sub-dominant
contribution in the nucleus.

The broad spectral coverage of our data allows us to determine the
continuum level in regions of the spectrum that are expected to be
free of line emission from any CO that might be present. After
continuum subtraction, we are left with a few potential emission
features (in 100~$\rm km\:s^{-1}$ channels) that are at strengths of
about 3-4 times the continuum fluctuations in the line-free
regions. These features may be due to systematic errors associated
with the continuum subtraction. Alternatively, if interpreted as CO
emission, the implied gas mass is about $5\times 10^6\:M_\odot$. Our
conservative estimate for the upper mass limit of molecular gas that
is present in a 700~$\rm km\:s^{-1}$ wide velocity range is also at
about this level, i.e. $8\times 10^6\:M_\odot$. Ignoring systematic
errors, as is typically done in such estimates, we have a $3\sigma$
mass sensitivity to about $3\times 10^6\:M_\odot$.


The presence of molecular gas inside the Bondi radius of M87's central
black hole is a strong discrimator of different accretion models. If
present it would indicate that gas has been able to cool to low
($T\sim 100$~K) temperatures, presumably in the confines of a thin
accretion disk, whose presence is already indicated by the
observations of atomic lines (Macchetto et al. 1997). This structure
is not part of the paradigm of RIAF models that are fed by Bondi
accretion (e.g. Di~Matteo et al. 2003). Although a RIAF could exist at
the center of the disk, much closer to the black hole, the accretion
flow at its outer boundary would only be that fraction of the mass
flux making it through a potentially star-forming disk, not the
unimpeded Bondi accretion rate.

The presence of molecular gas is a prediction of the star-forming
accretion disk model of TB05, in which strong gravitational
instability in the disk causes most of the accreted mass to form stars
at relatively low-luminosity rather than join the black hole. The
predicted molecular gas mass is $(1-5) \times 10^6 M_\odot$, which,
while quite uncertain, is consistent with the observational
constraints presented here. Higher sensitivity observations are
required to place more stringent constraints on the presence or not of
molecular gas in the circumnuclear disk of M87.





\acknowledgements We thank Melanie Krips, Alister Graham, Jonathan
Braine and Paola Caselli for discussions and the helpful comments of
the anonymous referees. JCT acknowledges support from CLAS, Univ. of
Florida and NSF CAREER grant AST-0645412.  HB~acknowledges financial
support by the Emmy-Noether-Program of the Deutsche
Forschungsgemeinschaft (DFG, grant BE2578); EGB acknowledges support
from NSF grants AST-0406799, AST-0406823, and NASA grant
ATP04-0000-0016 (NNG05GH61G).


\end{document}